\documentclass[aps, 
 amsmath,amssymb, amsbsy,
pre,
floatfix,
superscriptaddress,
twocolumn
]{revtex4-2}

\usepackage{gensymb}
\usepackage[caption=false]{subfig}
\usepackage{xcolor}
\usepackage{graphicx}
\usepackage{dcolumn}
\usepackage{bm}
\usepackage{float}
\usepackage{hyperref}
\usepackage{enumerate}
\usepackage{dsfont}
\usepackage{textcomp}
\usepackage{verbatim}
\usepackage{booktabs}
\usepackage{multirow}

\usepackage{calligra} 
\DeclareMathAlphabet{\mathcalligra}{T1}{calligra}{m}{n} \DeclareFontShape{T1}{calligra}{m}{n}{<->s*[2.2]callig15}{} 

\begin{document}

\title{Prospecting for Pluripotency in Metamaterial Design}

\author{Ryan van Mastrigt}
\email{r.vanmastrigt@uva.nl}
\affiliation{Institute of Physics, Universiteit van Amsterdam, Science Park 904, 1098 XH Amsterdam, The Netherlands}
\affiliation{AMOLF, Science Park 104, 1098 XG Amsterdam, The Netherlands}

\author{Marjolein Dijkstra}
\affiliation{Soft Condensed Matter \& Biophysics, Debye Institute for Nanomaterials Science, Department of Physics, Utrecht University,
Princetonplein 5, 3584 CC Utrecht, The Netherlands}

\author{Martin van Hecke}
\affiliation{AMOLF, Science Park 104, 1098 XG Amsterdam, The Netherlands}
\affiliation{Huygens-Kamerling Onnes Lab, Universiteit Leiden, Postbus 9504, 2300 RA Leiden, The Netherlands}

\author{Corentin Coulais}%
\affiliation{Institute of Physics, Universiteit van Amsterdam, Science Park 904, 1098 XH Amsterdam, The Netherlands}


\date{\today}

\begin{abstract}{
From self-assembly and protein folding to combinatorial metamaterials, a key challenge in material design is finding the right combination of interacting building blocks that yield targeted properties.
Such structures are fiendishly difficult to find---not only are they rare, but often the design space is so rough that 
gradients are useless and
direct optimization is hopeless.
Here, we design ultra rare combinatorial metamaterials capable of multiple desired deformations by introducing a two-fold strategy that avoids the drawbacks of direct optimization.
%
We first combine convolutional neural networks with genetic algorithms to 
prospect for metamaterial designs with a potential for high performance; in our case, these metamaterials have a high number of spatially extended modes---they are pluripotent.
%
Second, we exploit this library of pluripotent designs to generate metamaterials with multiple target deformations, which we finally refine by strategically placing defects.
%
Our pluripotent, multishape metamaterials would be impossible to design through trial-and-error or standard optimization. Instead, our data-driven approach is systematic and ideally suited to tackling the large and intractable combinatorial problems that are pervasive in material science.
%
}
\end{abstract}

\maketitle

Pluripotent mechanical metamaterials support a small number of spatially textured soft modes,  enabling exceptional functionalities such as selective mechanical responses~\cite{rocklin2017transformable, kim2019conformational, bossart2021oligomodal, meng2022deployable, mastrigt2023emergent, dykstra2023inverse}, nonlocal resonances~\cite{bossart2023extreme}, multi-shape folding~\cite{cho2014engineering, sussman2015algorithmic, overvelde2016three, overvelde2017rational, dieleman2020jigsaw, mcinerney2020hidden, melancon2022inflatable, aksoy2022multistable}
, and sequential energy-absorption~\cite{liu2023leveraging}.
To achieve such metamaterials, combinatorial strategies based on a small number of building blocks
are ideal~\cite{coulais2016combinatorial}.
Designing combinatorial metamaterials requires identifying intricate combinations of  building blocks that support targeted soft modes.
Because soft modes are highly sensitive to minute changes in these combinations, 
the design space is rough:
targeted designs are rare exceptions in a sea of random, failed designs~\cite{mastrigt2022machine}. Hence, while 
problems with continuum design spaces allow for  direct, gradient-based
optimization of an objective function~\cite{kim2019conformational, guo2020semi, mao2020designing, zheng2021data, bastek2022inverting, telgen2022topology, zheng2023deep, brown2023deep, ha2023rapid}, such single-step optimization fails for combinatorial metamaterials. As a result, the systematic design of pluripotent metamaterials 
remains an open problem.

\begin{figure}[t!]
    \centering
    \includegraphics[width=\columnwidth]{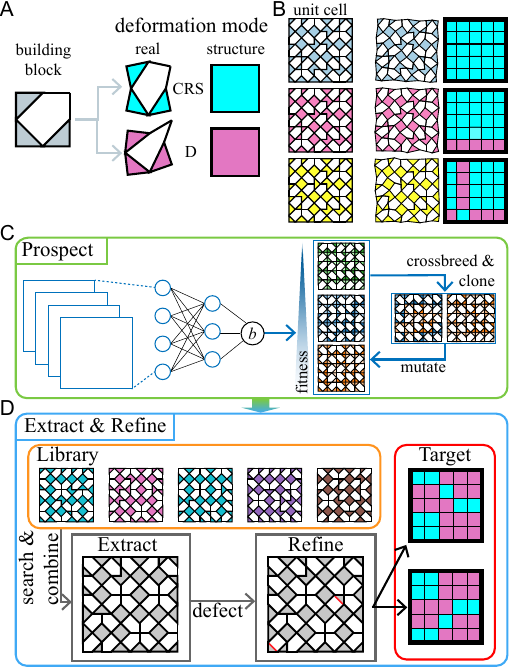}
    \caption{
    Design approach for combinatorial metamaterials.
    \textbf{(A)} The two-dimensional building block (left, gray) can be tiled in four orientations and features two distinct zero modes (middle), which we label CRS (top, cyan) and D (bottom, pink). We visualize the mode structure of a deformed building block as a cyan or pink square (right).
    \textbf{(B)} The building blocks of (A) combine into larger $5\times 5$ unit cells (left) that feature intensive zero modes (middle). The mode structure of such intensive modes is shown on the right. 
    The three unit cells (blue, pink, yellow) differ only by a single building block from their neighbors, yet support an increasing number of modes:
        each unit cell supports the zero modes of their top neighbors in addition to the zero mode directly to the right of the unit cell. 
    \textbf{(C)} Step (i) of our design approach uses a convolutional neural network (CNN) that predicts the number of intensive modes $b$ to guide a genetic search algorithm (GA) to efficiently generate high $b$ (high pluripotency) designs.
    \textbf{(D)} In step (ii), the generated high-pluripotency designs are added to a library. We then search and combine designs from this library to find a design that closely matches a set of target deformations (red): this process is called  extraction. Finally,  our extracted designs often require further refinement which we 
    implement by strategically introducing  
    defects. The final refined design features the desired target deformation modes while minimizing  undesired superfluous modes.
    }
    \label{fig: design space}
\end{figure}

Here, we introduce a data-driven 
design approach to find rare combinatorial designs without relying on gradients. 
We consider a family of mechanical metamaterials with multiple soft modes [Fig.~\ref{fig: design space}(A)-(B)]. This problem is emblematic: the design space is extremely rugged, explicit design rules are unavailable, and evaluating the modes
for each design is computationally heavy.
To find rare designs with desired modes, we propose to initially  search for designs with high \emph{pluripotency}, rather than  high fitness. 
In other words, instead of directly attempting to design metamaterials with a few target modes, we first construct a library of designs that have the potential to exhibit as many target modes as possible---they are pluripotent. In a second step, we use this library to generate designs. This approach allows us to computationally find metamaterials that host target modes with  specific deformations that would otherwise be impossible to design.

To illustrate our proposed approach, we draw an analogy to the process of finding, extracting, and refining rare metals: we first prospect for ore, and subsequently extract to obtain gold.
Because we initially focus  on  general pluripotency rather than specific fitness, our design approach consists of two steps: (i) prospecting for a library of high pluripotency candidate designs using a genetic algorithm (GA) guided by a convolutional neural network (CNN) [Fig.~\ref{fig: design space}(C)] and (ii) \emph{extracting} and \emph{refining} candidate designs to find a design with targeted properties [Fig.~\ref{fig: design space}(D)]. 


Our two-step approach presents a new avenue for the systematic design of pluripotent (meta)materials. While so far metamaterial design was limited to a single target mode and known design rules
~\cite{coulais2016combinatorial, dieleman2020jigsaw, pisanty2021putting}, our data-driven approach allows us to design 
pluripotent metamaterials with multiple target modes
without the necessity of knowing the design rules.
Specifically, we will demonstrate that our approach allows us to design large metamaterials with two desired textured soft modes, e.g., that resemble a smiley and frowny face, or the letters A and U, while minimizing superfluous modes.
Beyond metamaterial design, we anticipate  that our approach will be applicable to a wide range of combinatorial problems characterized by multiple, independent sets of constraints to satisfy. Such problems can readily be found in fields such as protein folding~\cite{huang2016coming, lovelock2022road}, self-assembly~\cite{zeravcic2014size, evans2024pattern}, computer graphics~\cite{merrell2007example, merrell2010model}, and molecular design~\cite{bilodeau2022generative}.

\section*{Multimodal metamaterial and design strategy}

\begin{figure*}[t!]
    \centering
    \includegraphics{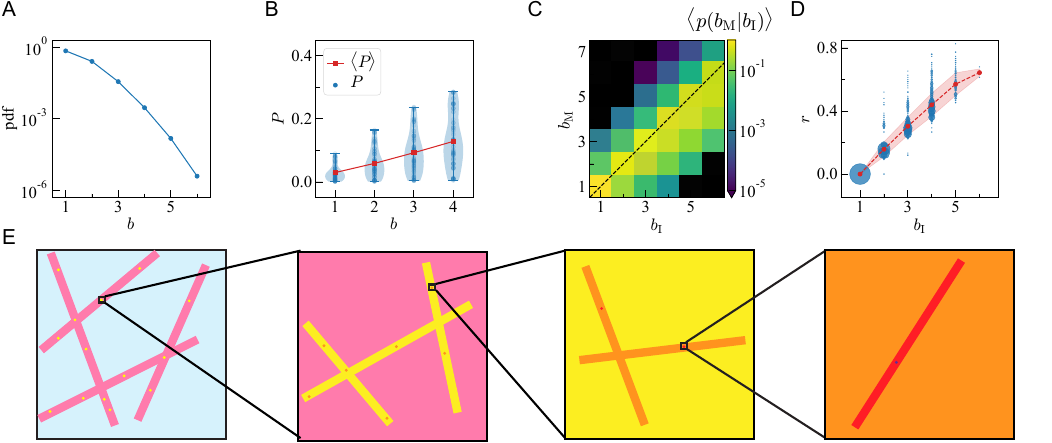}
    \caption{Statistical  characterization of the design space. 
    \textbf{(A)} Probability density function (pdf) for the number of intensive modes $b$ obtained through Monte Carlo sampling of $5\times 5$ unit cells. 
    \textbf{(B)} An increasing number of intensive modes $b$ correlates to a higher average capacity [\eqref{eq: C}] or pluripotency $P$ per design (blue circles). The pluripotency averaged over random designs (red squares) increases with $b$. The blue shaded area represents a violin plot to visualize the distribution of $P$ values.
    \textbf{(C)} The average probability $\langle p (b_{\mathrm{M}} | b_{\mathrm{I}})\rangle$ of finding a mutated unit cell with $b_{\mathrm{M}}$ intensive modes under point mutations for an initial unit cell with $b_{\mathrm{I}}$ intensive modes. To highlight the low probability of transitioning to a higher $b$,  a logarithmic colormap (colorbar) is used, with zero (measured) probability denoted in black. 
    \textbf{(D)} The average ratio $r$ (red dashed line, shaded area indicates the standard deviation) of mutated unit cells with $b_{\mathrm{M}} \leq b_{\mathrm{I}}$ averaged over random unit cells with $b_{\mathrm{I}}$ intensive modes increases with $b_{\mathrm{I}}$. The local $r$ varies for individual unit cells (blue circles, where the size indicates the number of unit cells with the same $r$) and appears multimodal in structure. This is likely a consequence of the different types of intensive modes: edge and global.
     \textbf{(E)} Conceptual configuration space of a discrete combinatorial multimodal metamaterial problem. Areas of the same color represent designs with the same $b$, where blue, pink, yellow, orange, and red correspond to areas with progressively increasing $b$. Random point-mutations of the design---rotating a single building block---generally lower the number of intensive modes $b$, most designs have a low $b$ and we have few examples of rare designs with higher values of $b$. Here, satisfying a hierarchy of conditions leads to increasingly low-dimensional subspaces where $b$ has larger values (left to right).
    }
    \label{fig:b characterization}
\end{figure*}

To test our design approach, we consider the spatial structure of zero modes---infinitesimal deformations that do not stretch any  bonds  to first order---in 
a multimodal combinatorial metamaterial [Fig.~\ref{fig: design space}(A)-(B)]~\cite{bossart2021oligomodal, mastrigt2022machine, mastrigt2023emergent}.
The orientation of each building block 
specifies the number and spatial structure of the zero modes, leading to a rough design space.
While discrete tiling based techniques have been developed for combinatorial metamaterials that support a single spatially textured zero mode~\cite{coulais2016combinatorial, meeussen2020topological, pisanty2021putting}, the task of designing
pluripotent metamaterials 
that feature multiple target modes is 
significantly harder and
systematic techniques are lacking.

As a first step, we aim to 
design pluripotent metamaterials 
that feature many modes.
%
For our metamaterials, the number of zero modes, $N_M(n)$,
for an $n\times n$ tiling of unit cells (each containing $k\times k$ building blocks) grows with $n$ as:~\cite{mastrigt2022machine} 
\begin{equation}
    N_M = a n + b~.
    \label{eq: M(n)}
\end{equation}
We previously investigated the detection of rare unit cells with $a=1$ using CNNs, 
showed that these unit cells contain specific `strip' modes, and investigated their design and properties 
~\cite{mastrigt2022machine,liu2023leveraging,mastrigt2023emergent}. 
Here we focus on rare pluripotent designs with extended zero modes, which requires many 
spatially extended, intensive modes (large $b$) and no strip modes ($a=0$). 
For all designs, $b\geq 1$ due to the presence of the  well-known
 counter-rotating squares (CRS) mode~\cite{resch1965geometrical, evans2000auxetic, mullin2007pattern, bertoldi2010negative, coulais2015discontinuous, bertoldi2017flexible, coulais2018characteristic, bossart2021oligomodal, czajkowski2022conformal} [Fig.~\ref{fig:b characterization}(A)]. Yet, there are no design rules for realizing larger values of $b$, illustrating the need for
the development of an effective  search algorithm.

We use the deformation in each building block to 
characterize the structure of a zero mode $M$.
In particular, we decompose the 
deformations of 
the constituent building blocks $m_i$ into the trivial CRS deformation $m_{CRS}$ and the non-trivial deformation $m_{D}(c_i)$,
where $c_i$ denotes the orientation of building block $i$:
$m_i = \alpha_i ~ m_{CRS} + \beta_i ~ m_{D}(c_i)$.
We label building block deformations with $\beta=0$ as CRS blocks and those with $\beta \neq 0$ as D blocks [Fig.~\ref{fig: design space}(A)]. We characterize the structure of  zero modes by the spatial distribution of CRS and D blocks,
and define a set of target modes 
$\{\hat{M}\}$ by its CRS and D blocks
[Fig.~\ref{fig: design space}(B)]~\cite{mastrigt2023emergent}. 

As the design space is rough, local modifications do not provide evidence whether a successful design is near, and 
finding a successful design by incremental modifications is not feasible. Our two-step strategy is to, first,  create a library of `high-potential', pluripotent designs, which is independent of the specific target modes; second, use this library to construct a design for the specific target modes.

Quantifying the pluripotency of a given $k\times k$ design 
involves considering both a set of zero modes of the material
$\{M\}$
and a set of targeted zero modes $\{\hat{M}\}$.
We define the similarity 
between  modes $M$ and  $\hat{M}$
 as  the fraction of matching CRS and D blocks:
\begin{equation}
    S(M, \hat{M}) = \frac{1}{k^2} \sum_{i=1}^{k^2}\begin{cases}
            \delta(\beta_i, 0), & \mathrm{if} \, \hat{\beta}_i = 0 \\
            1-\delta(\beta_i, 0), & \mathrm{if} \, \hat{\beta}_i \neq 0
    \end{cases}
     \label{eq: C}
\end{equation}
where $\delta$ represents the Kronecker delta.
We then define the capacity of a design as the maximum similarity between the target mode and  linear combinations of the zero modes $\{M\}$ of the design:
\begin{equation}
    C(\{M\}, \hat{M}) = \mathrm{max}_{\mathbf{w}}\left[S(\sum_i w_i M_i, \hat{M})\right] - \hat{N}_{\mathrm{CRS}},
    \label{eq: C set of M}
\end{equation}
where $w_i$ are weights and $\hat{N}_{\mathrm{CRS}}$ 
is the fraction of CRS blocks in the target mode. We calculate the maximum similarity  
using constraint programming (see Materials and Methods). We finally
define the pluripotency $P$ of a design as the average capacity $C$ computed over a randomized set of target modes (see Materials and Methods).

Of course, the capacity is maximal for extremely floppy designs,
but their excess floppy modes hinder their functionality ~\cite{dykstra2023inverse}.
Ultimately, we aim to construct metamaterials which maximize $C$ without any excess floppy modes, and
our oligomodal combinatorial metamaterials are ideally suited to do so \cite{bossart2021oligomodal}.
We now show that designs with a larger number of 
intensive modes $b$ result in higher pluripotencies.
We first generate a random set of 100 target modes 
(See SI). We then generate $10^6$ random designs, calculate $b$ for each, and then calculate their pluripotency 
(see methods).  
Thus, we take $b$ as a proxy for pluripotency $P$.

\section*{High potency designs}
We now prospect for high pluripotency by searching for designs with values of $b$ that are larger than can be obtained by random sampling. While we have indications for their existence, we 
lack examples of such designs. 

\begin{table}[b!]
    \centering
    \caption{\label{tab: CM b} Confusion matrix for the test set of the CNN with the lowest validation loss.}
    \begin{tabular}{*{7}{c}}
    &&  \multicolumn{5}{c}{predicted $b_{CNN}$}\\\midrule
       & & 1 & 2 & 3 & 4 & 5 \\
      \multirow{5}{*}{actual $b$} & 1 & 105767 & 47 & 0 & 0 & 0 \\
      & 2 & 418 & 37877 & 38 & 0 & 0 \\
      & 3 & 1 & 242 & 5101 & 16 & 0 \\
      & 4 & 0 & 0 & 66 & 383 & 3 \\
      & 5 & 0 & 0 & 1& 5 & 17 \\
      \bottomrule
    \end{tabular}
\end{table}

We first explore the design space of $5\times 5$ unit cells, as this size offers a good balance between spatial complexity, the size of the design space, and the rarity of the number of intensive modes $b$. 
We note that designs with large values of $b$ are
increasingly rare [Fig.~\ref{fig:b characterization}(A)], yet those are interesting because they tend to be pluripotent [Fig.~\ref{fig:b characterization}(B)].
Crucially, there is structure within the design space, 
which we explore by characterizing the changes in $b$ under point mutations of the design, defined as a 
random rotation of one of the building blocks [Fig.~\ref{fig:b characterization}(C)]. 
Comparing the number of intensive modes of the initial 
design, $b_{\mathrm{I}}$, to the number of modes of the mutated design, $b_{\mathrm{M}}$ 
we find that the most likely scenario is that 
$b_{\mathrm{M}}=b_{\mathrm{I}}$, showing that designs with the same value of $b$ are interconnected. The next most likely scenario is that $b_{\mathrm{M}}=b_{\mathrm{I}}-1$, showing that designs with $b$ zero modes are surrounded by designs with $b-1$ zero modes.
The rarest yet desired scenario is that $b_{\mathrm{M}}=b_{\mathrm{I}}+1$.
To gain further insight, we explored the ratio $r$ of neighbors that have $b_{\mathrm{M}}<b_{\mathrm{I}}$, and found that this ratio increases with $b_{\mathrm{I}}$, showing that subspaces with large $b$ have increasingly small dimensionality [Fig.~\ref{fig:b characterization}(D)].
This means that the subspaces of the design space with constant $b$ become progressively sparse and low-dimensional with increasing $b$. 
We describe this hierarchical structure as `needles-within-needles-within-needles in a haystack' [Fig.~\ref{fig:b characterization}(E)], which stands in stark contrast to the
less structured,
jagged structure of the direct objective function in design space. 

We now turn our attention to a computationally effective method to estimate $b$. A straight-forward calculation of $b$ requires determining the number of zero modes as 
function of the number of unit cells $n$,
which is computationally demanding ~\cite{hong1992rank, gu1996efficient}. 
Inspired by the earlier success of 
convolutional neural networks (CNNs) in classifying designs with $a\geq1$, we now set out to use CNNs to estimate $b$~\cite{mastrigt2022machine}. We train our CNN using 
training data obtained by Monte Carlo sampling of the design space for $5\times 5$ unit cells and calculating $b$ from the values of $N_M(n)$ for $n\in \{2, 3, 4\}$ 
(see Materials and Methods). Despite the sparsity of designs with a high number of intensive modes $b$ in the training set, we find that the CNN remains accurate for high $b$ in the test set (Tab.~\ref{tab: CM b}). In particular, the output of the CNN, $b_{CNN}$, is consistently close to the actual number of intensive modes $b$, albeit with a slight underestimation bias. Crucially, our 
CNN is two orders of magnitude faster and is readily parallelizable~\cite{mastrigt2022machine}.
Thus, the CNN is able to quickly and accurately detect the needles for low $b$; whether it can identify ultra rare designs with large $b$ remains an open question at this point.

\begin{figure}[t]
    \centering
    \includegraphics[width=\columnwidth]{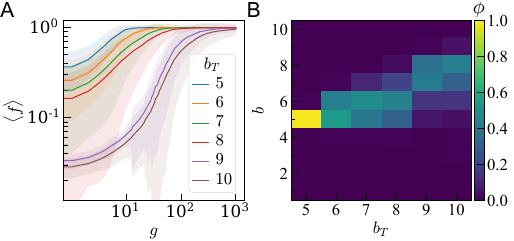}
    \caption{Extrapolation of the trained CNN. 
    (\textbf{A}) The average (solid line) and standard deviation (shaded area) of the fitness $f$ [\eqref{eq: fitness}] of the fittest design per generation $g$ over two hundred GA runs for each target number of intensive modes $b_T$ (legend). 
    (\textbf{B}) Fraction $\phi(b)$ of the fittest designs with $b$ intensive modes for a given target $b_T$ across two hundred GA runs per $b_T$. 
    }
    \label{fig:GA results}
\end{figure}

Finally, we now employ a combination of a genetic search algorithm (GA) with the trained CNN to 
iteratively progress towards designs with a high number of intensive modes $b$ (see Materials and Methods).
Specifically, we use the GA to find designs with a target number of intensive modes $b_T$ by maximizing the fitness
\begin{equation}
    f = \frac{1}{1 + (b_{CNN}-b_{T})^2}~.
    \label{eq: fitness}
\end{equation}
Surprisingly, starting from a Monte Carlo sampled set of unit cell designs, the GA consistently reaches its maximum fitness relatively quickly, 
even when $b_T$ exceeds the range of the training data, i.e., $b_T > 6$ [Fig.~\ref{fig:GA results}(A)]. 
Even though the CNN overestimates the number of intensive modes, a significant fraction of these designs indeed features $b>6$ [Fig.~\ref{fig:GA results}(B)]. 

Hence, our combination of GA and CNN is able to identify
extremely rare designs with high pluripotency (e.g., up to $b=9$). We estimate that such designs represent only a fraction of $\mathcal{O}(10^{-8})$ of the total design space; yet combining GA and CNNs allows us to find these within minutes on a desktop computer. We believe that the `needles-within-needles' structure of the design space is crucial. First, it allows the GA, by
a combination of local and nonlocal exploration, to 
incrementally increase $b$ (see Supplemental Information); second, we believe that this underlying structure allows the CNN to extrapolate beyond its training data. 
Hence, we have carried out the first step of our design strategy, by optimizing a
quantity---pluripotency---that has structure in the design space,
rather than directly optimizing an essentially random objective function.

\section*{Designing for target deformations \label{sec: target deformations}}

We now transform our candidate designs, which have a high pluripotency based on their capacity for 
random targets, 
into  a specific design with a high capacity for specific targets $\{\hat{M}\}$. This is 
step (ii) of our design strategy, which itself proceeds in two substeps, which we refer to as 
extract and refine. 

\begin{figure*}[t]
    \centering
    \includegraphics[width=\textwidth]{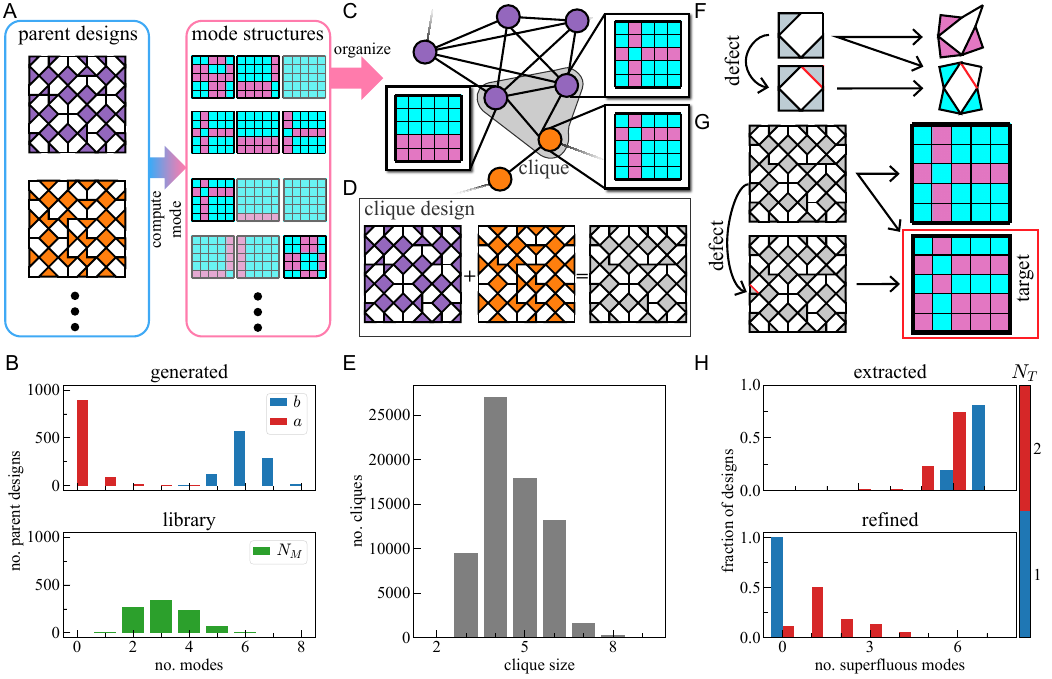}
    \caption{
    Extracting and refining high pluripotency designs. 
    (\textbf{A}) We use 1000 GA-generated designs (left), calculate their mode structures (right), and discard the common CRS mode (all blue) and simple edge modes (single-block-wide strip(s) of D blocks located at the edge), as indicated by the grayed-out mode structures. The remaining modes are organized in a graph.
    (\textbf{B}) Top: histogram of the true number of intensive modes $b$ (blue left bars) and extensive modes $a$ (red right bars) for 1000 GA-generated parent designs with a target of $b_T=7$. Bottom: histogram of the number of modes $N_M$ (green bars) after discarding the common CRS mode and edge modes (see Materials and Methods) per parent design in the library. The total number of modes is significantly reduced after this discarding.
    (\textbf{C}) Individual modes from distinct parent designs (indicated by color) are represented as nodes in a graph. Nodes are connected by edges if the corresponding modes are compatible. Each maximal clique (fully-connected nodes enclosed in gray area) corresponds to a new design [see (D)] that supports all constituent modes. 
    (\textbf{D}) The clique in (C) 
    comprises modes from two parent designs (purple and orange) which we combine into a new design (gray) that supports all three modes in the clique. 
    (\textbf{E}) Histogram of maximal clique sizes (number of nodes) in the library. We note that a larger clique size corresponds to a higher number of modes and thus a higher (average) pluripotency $P$ [Fig.~\ref{fig:b characterization}(B)]. 
    (\textbf{F}) Introducing a defect (extra rigid bar, red) to a building block (bottom left) prohibits the D mode (top right, pink) while retaining the CRS mode (bottom right, cyan). Arrows indicate supported modes. 
    (\textbf{G}) Strategic placement of a defect (red) prohibits an undesired mode structure (top right) while retaining the desired target mode structure (bottom right).
    (\textbf{H}) Normalized histograms 
    of the number of superfluous modes\ref{fn: CRS neglect} for 100 designs found in our library with the highest cumulative capacity [\eqref{eq: C set of M}] for $N_T$ (colorbar) randomized target deformations before (top) and after (bottom) strategically adding defects (refining). We neglect the trivial global CRS mode, as we are free to remove this mode by adding a single extra horizontal or vertical rigid bar. 
    }
    \label{fig:library_refining}
\end{figure*}

In the extracting step, we first generate a library of 1000 pluripotent parent designs and use these to generate a much larger space of designs
[Fig.~\ref{fig:library_refining}(A)]. We set the target for the number of intensive modes  for the parent designs as $b_T=7$, and find that the distribution of their actual values of $b$ is centered around $b=6$ [Fig.~\ref{fig:library_refining}(B)].
Next, we compute all corresponding zero modes,
remove the trivial CRS mode and most edge modes, thus obtaining
the spatially extended zero modes of each design
[Fig.~\ref{fig:library_refining}(A)].

We then generate a very large number of  high potency designs (`children') as follows. We define two modes as compatible when
they have no D blocks with different orientations at the same site [Fig.~\ref{fig:library_refining}(C)].
We construct a graph where the nodes represent the modes and the edges indicate pairs of compatible modes.  
We then determine the maximal
cliques---sets of nodes such that every two constituent nodes are adjacent (clique) that cannot be extended by adding additional nodes (maximal)---of this graph using the Bron-Kerbosch algorithm~\cite{bron1973algorithm}. Cliques correspond to sets of compatible zero modes [Fig.~\ref{fig:library_refining}(C)]. 
Crucially,
while each parent design corresponds to a clique, the large number of additional maximal cliques stem from multiple parent designs. For each of these, we can rationally combine the relevant parent designs to create a new design (`child') that precisely generates the additional clique (see Materials and Methods) [Fig.~\ref{fig:library_refining}(D)]. 
These additional designs significantly expand the search space: the number of maximal cliques exceeds the number of initial designs by a factor 70, and many of these correspond to
designs with a high clique size and thus a corresponding high pluripotency [Fig.~\ref{fig:library_refining} (E)].
Then, to extract a design 
we select the maximal clique with the highest capacity $C$ [\eqref{eq: C set of M}] with respect to the target modes $\{\hat{M}\}$,
and consolidate the constituent designs into a new, highest-capacity, single candidate design that supports all modes in the clique [Fig.~\ref{fig:library_refining}(D)].

To test the capabilities of this large set of designs, we define 100 random targets and study the success rate of finding a design
that contains a set of modes that  perfectly match a target ($C=1$). 
For single randomized target deformations 
we successfully find a
design in all cases; for two randomized target deformations we find designs in 82 
cases (see Materials and Methods). Hence, extending the design space with maximal cliques allows us to find designs that satisfy multiple target deformations  with a high probability.

Second, we refine. We note that most of our designs originate from large cliques which feature many modes, most of which are superfluous. Crucially, removing undesired modes is much easier than adding desired modes: a mode requires a careful selection of building blocks to deform compatibly, while changing a single building block can be sufficient to prohibit the mode. We thus eliminate superfluous modes by introducing point defects in our design [Fig.~\ref{fig:library_refining}(F)-(G)].
Specifically, by adding an extra diagonal bar, each building block can be prevented from deforming in the D mode  and can only deform as a CRS block [Fig.~\ref{fig:library_refining}(F)].
We iterate this procedure until we can no longer  transform D blocks to CRS blocks, or until only the desired modes are left. This approach drastically reduces the number of superfluous modes [Fig.~\ref{fig:library_refining}(H)].

Together, this two-step strategy allows us to 
obtain targeted pluripotent designs with few superfluous modes: 
For a single target deformation mode, we are able to completely eliminate any superfluous modes\footnote{we neglect the trivial global CRS mode that is always present, as we can remove this mode while retaining all other modes by adding a single extra horizontal or vertical rigid bar to the design \label{fn: CRS neglect}}. For two target deformations, most refined designs feature a single superfluous mode, and the number of designs with a higher number of intensive modes rapidly tapers off.
Thus, we are able to find extremely rare designs that exhibit target deformations while minimizing the number of superfluous deformations.

\section*{Scaling up}
We now show how we can extend our approach to obtain larger designs by stitching together $5\times 5$ designs. 
Specifically, we create two $10\times 10$ metamaterials that each support two zero modes with spatially complex structures: one resembling a smiley and frowny face, and another resembling the letters A and U.

\begin{figure*}[t!]
    \centering
    \includegraphics[width=\textwidth]{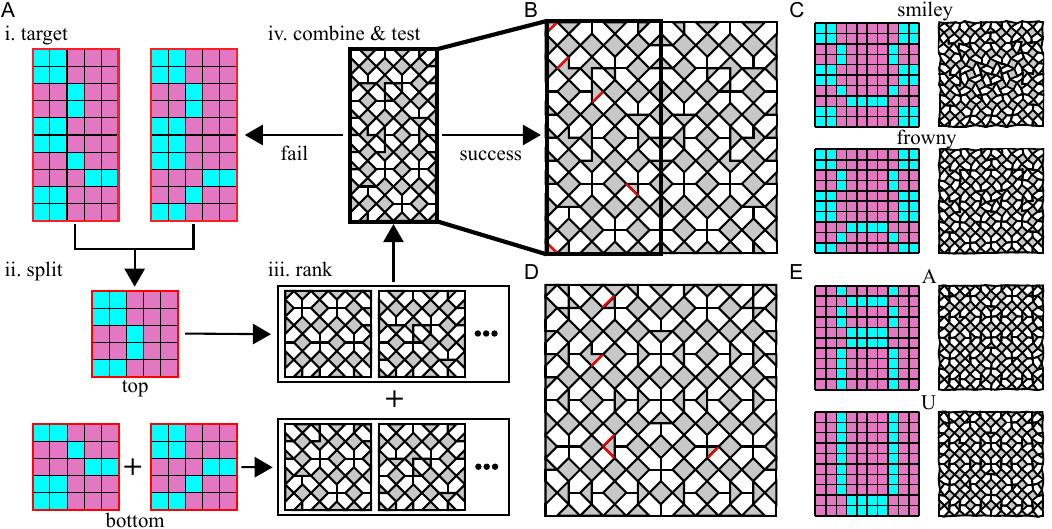}
    \caption{Combining designs for larger target deformations.
    (\textbf{A}) Strategy for combining designs in four substeps. In step i, the two target modes (highlighted with a red border) are split (step ii) into separate sets of target modes for the top and bottom $5\times 5$ designs. In step iii, the maximal cliques are ranked by the average capacity [\eqref{eq: C set of M}] for each set of target modes.
    In step iv, we progress iteratively through combinations of the highest-ranking designs until a combination with a sufficiently high capacity with respect to the two target modes of step (i) is found. This final design is mirrored and refined to form the final $10\times 10$ design of (B).
    (\textbf{B}) The refined (defects are highlighted in red) $10\times 10$ design for the target smiley and frowny zero modes.
    (\textbf{C}) The design of (B) supports two zero modes with structures resembling a smiley and frowny face. 
    (\textbf{D}) Refined $10\times 10$ design for the target A and U zero modes. 
    (\textbf{E}) The design of (D) supports two deformation modes with structures resembling the letters A and U.}
    \label{fig:smiley experimental}
\end{figure*}

We demonstrate this strategy first for the smiley/frowny design target.
We start by using symmetries to reduce the two targeted $10\times 10$ mode structures into two $5\times 10$ target modes (Fig.~\ref{fig:smiley experimental}(A)-i); these can be split into a $5\times5$ top  with one zero mode, and a 
$5\times5$ bottom with two zero modes (Fig.~\ref{fig:smiley experimental}(A)-ii). 
As before, for each of these, we determine and rank all maximal cliques (Fig.~\ref{fig:smiley experimental}(A)-iii). 
The non-trival step is to combine the top and bottom: in general, combining two unit cells prohibits certain modes due to incompatible kinematic constraints at the interface. 
Therefore, we iterate over the next-highest ranked combinations if the combined design is not satisfactory (Fig.~\ref{fig:smiley experimental}(A)-iv). Using this approach, we find a successful $5\times 10$ design; and using mirror symmetry, this yields 
a $10\times 10$ design that features a mode structure resembling a smiley and frowny face [Fig.~\ref{fig:smiley experimental}(B-C)].

In addition to the smiley and frowny modes, we find seven superfluous modes\ref{fn: CRS neglect}. 
We refine this design by strategically adding five extra rigid bars which reduces the number of superfluous modes to two [Fig.~\ref{fig:library_refining}(G)]. 
We suggest that this design is an excellent starting point for further refinement to eliminate all spurious modes, e.g., by replacing a few building blocks with newly designed building blocks that only feature a single, complex deformation.

We follow the same procedure to find a design that features modes whose structures resemble the letters A and U, spelling Au for gold. Our best design features 11 superfluous modes, in addition to the two desired modes; by adding 
five rigid bars the number of superfluous modes is reduced to six [Fig.~\ref{fig:smiley experimental}(E)].  Thus, our design strategy allows us to find extremely rare designs with multiple desired modes within an otherwise intractable design space.

\section*{Discussion}
Combining building blocks to create structures with desired properties is notoriously difficult---the problem is often ill-posed, and the design space is too vast to fully explore. Without access to underlying design rules, directly navigating such spaces with a strongly discontinuous objective function to find designs with desired properties is hopeless.
In this paper, we introduced a general strategy to find  ultra rare designs using pluripotency: a statistical measure that quantifies performance over a class of problems.
In short, our approach exploits the  hierarchical structure of pluripotency in the design space to generate a library of many high pluripotency designs. Subsequently, we select and refine from this library to reach the final, ultra rare design that satisfies the desired properties.

Our approach opens up new exciting avenues for combinatorial design. For example, our approach can be readily applied to metamaterial design for a plethora of different building block designs, allowing for much faster exploration of the vast, largely unexplored design space of metamaterial geometries. Moreover, we foresee applications beyond the field of metamaterials. Self-assembling systems, for example, require the right set of building blocks to achieve the desired end geometry---which is hard without access to assembly rules~\cite{hormoz2011design, zeravcic2014size, gartner2022time}. Additionally, information processing in designer matter can be described by a set of hysteretic elements---how to order and tune interactions between these elements to achieve complex memory properties is an open challenge~\cite{hecke2021profusion, bense2021complex, shohat2022memory}.

\textit{Data availability statement.}\textemdash
The codes to calculate zero modes and their structures~\cite{MastrigtGitLabZM}, and to design~\cite{MastrigtGitLabID} are freely available. Additionally, the data used to train our neural networks, and all our generated and refined designs are also freely available~\cite{Zenodo_ID}.


\textit{Acknowledgments.}\textemdash
This work was carried out on the Dutch national e-infrastructure with the support of SURF Cooperative. C.C. acknowledges funding from the European Research Council under grant no. ERC-2019-STG 852587 and from the Netherlands Organization for Scientific Research under grant agreement VIDI 213131. M.D.
acknowledges financial support from the European Research Council under grant no. ERC-2019-ADV-H2020 884902.


\appendix
\section{Materials and Methods}
\subsection{Calculating the number of zero modes}
We determine the number of zero modes $N_M$ of a design in two steps. First, we construct the compatibility matrix $\mathcal{C}$ of the design's corresponding bar-joint framework. This matrix maps the deformations of the joints to elongations of the bars~\cite{calladine1978buckminster, lubensky2015phonons}. Zero modes are infinitesimal deformations of the joints that do not stretch any of the bonds up to first order. Mathematically, the dimension of the kernel or null-space of the compatibility matrix $\mathcal{C}$ corresponds to the number of zero modes. Thus, in our second step, we calculate this dimensionality using rank-revealing QR (rrQR) decomposition~\cite{hong1992rank, gu1996efficient}. The time complexity of this algorithm is $\mathcal{O}(n^3)$ for $n\times n$ matrices.

To calculate the coefficients $a$ and $b$ from our mode-scaling relation [\eqref{eq: M(n)}], we use rrQR to calculate the number of modes $N_M(n)$ for $n\times n$ tilings of unit cells with open boundary conditions in the range $n\in\{1, 2, 3, 4\}$. We use $N_M(3)$ and $N_M(4)$ to determine the slope $a$ and offset $b$. We use 1M $5\times 5$ unit cells drawn from an uniform discrete distribution to obtain the distribution of $b$ shown in Fig.~\ref{fig:b characterization}(A).


To obtain Fig.~\ref{fig:b characterization}(B)-(C), we selected at most 1000 designs per initial number of intensive modes $b_{\mathrm{I}}$ from the set of Monte Carlo sampled designs. For $b_{\mathrm{I}} = 5$ and $b_{\mathrm{I}}=6$, there are fewer than 1000 designs available, instead we use only 311 and 8 designs, respectively. For each of the selected designs, we infer the number of strip modes $a$ and intensive modes $b$ for the 75 designs that differ from the selected design by a single building block mutation. We determine the ratio of neighboring designs with $b_{\mathrm{NN}}$ intensive modes to the total number of neighboring designs for each design and average over all designs for a given initial number of intensive modes $b_{\mathrm{I}}$ to obtain the probability density function $p(b_{\mathrm{NN}} | b_{\mathrm{I}})$. Similarly, we define the normalized codimension for a design with $b_{\mathrm{I}}$ modes as the fraction of all neighboring designs with $b_{\mathrm{NN}}\leq b_{\mathrm{I}}$.

\subsection{Determining the structure of a zero mode}
The structure of zero modes can be defined in terms of individual building block deformations. As described in the main text, our building blocks feature two zero modes: the CRS mode $m_{CRS}$ and D mode $m_{D}$.
A building block can deform with any linear combination of these two independent modes: $m = \alpha m_{CRS} + \beta m_{D}$. We classify building blocks by these zero modes: if a block deforms with $\beta=0$, it is a CRS block, and if a block deforms with $\beta \neq 0$, it is a D block. In earlier work, we showed that there are limitations on the spatial structure of zero modes in tilings of these building blocks: regions of adjacent CRS blocks must always be rectangular of shape~\cite{mastrigt2023emergent}.

To determine the structure of zero modes supported by a given design in terms of CRS and D blocks, a simple calculation of the kernel of the compatibility matrix $\mathcal{C}$ is no longer sufficient. Instead, we directly solve for the kinematic degrees of freedom of the building block: $\alpha$ and $\beta$. These kinematic degrees of freedom effectively describe the infinitesimal change in each of the five free interior angles of the building block to first order. The kinematic constraints between neighboring building blocks can be described as constraints of adjacent angles of building blocks. For more technical details on this representation of modes, we refer to our paper~\cite{mastrigt2023emergent}. 

In short, we solve for each building block's kinematic constraints by composing a large integer matrix of all kinematic constraints between building blocks and use the Python package sympy~\cite{sympy} to find the null space of this matrix. This yields a set of rational vectors that form a basis for all valid zero modes in the design and translate directly to the kinematic degrees of freedom $\alpha$ and $\beta$. Thus, this method allows us to determine the structure of zero modes for a given metamaterial design in terms of CRS and D blocks. The code for this is freely available on our public GitLab~\cite{MastrigtGitLabZM}.


\subsection{Capacity of a set of modes \label{met: capacity}}
The capacity $C(\{M\}, \hat{M})$ [\eqref{eq: C set of M}] of a set of basis zero modes $\{M\}$ with respect to a target mode $\hat{M}$ is $S(M^*, \hat{M})$ [\eqref{eq: C}], where $M^*=\sum w_j M_j$ is a linear combination of the modes, and the set of weights $\mathbf{w}$ maximizes $S$. To find this set of weights, we use constraint programming~\cite{cpsatlp}. 

Specifically, we consider the D mode coefficients $\beta_{i}^j$ at site $i$ corresponding to the basis mode labeled by $j$. We define a set of integer variables $\mathbf{w}$ such that $M^*=\sum_j w_j M_j$. The goal is to find an $M^*$ that is close to the target mode $\hat{M}$. For every D site $i$ in the target mode, if any of the basis modes $\{M\}$ also has a D block at that site, we add a constraint $\sum_j w_j \beta^j_i \neq \delta_i$, where $\delta_i$ is a non-negative integer variable. Similarly, at every CRS site $i$ in the target mode, if any of the basis modes $\{M\}$ also has a D block at that site, we add the constraints $\sum_j w_j \beta_i^j \geq - \delta_i$ and $\sum_j w_j \beta_i^j \leq \delta_i$. The constraint programming solver then attempts to find a solution to all the variables that minimizes $\sum \delta_i$. The underlying idea is that by minimizing this sum, the program tries to find a set $\mathbf{w}$ for which most variables satisfy $\delta_i=0$. When $\delta_i = 0$ for all $i$, the mode $M^*$ has the same kind of block, CRS or D, at site $i$ as the target mode $\hat{M}$. The final similarity is then $S(M^*, \hat{M})$ [\eqref{eq: C}] with $M^* = \sum_j w_j M_j$. The code we used to calculate the capacity is publicly available~\cite{MastrigtGitLabID}.

\subsection{Randomized target deformations}
To determine the pluripotency $P$ of a design or a set of zero modes, we calculate the average  capacity $C$ [\eqref{eq: C set of M}] for a set of randomized target modes $\hat{M}$. We define our target modes $\hat{M}$ in terms of their structure: the spatial distribution of CRS and D blocks. However, our metamaterial has limitations regarding the  mode structures it can achieve. In previous work, we demonstrated that any valid mode must comprise  rectangular patches of adjacent CRS blocks~\cite{mastrigt2023emergent}. Thus, to assess our designs fairly, we cannot simply generate any random distribution of CRS and D blocks.

Instead, we generate viable target modes by generating a horizontal and vertical strip of D blocks within a background of CRS blocks. The position and width of these strips are randomly drawn from a uniform distribution. Where the two strips overlap, we replace the D blocks with  CRS blocks. This approach ensures that there are only rectangular patches of adjacent CRS blocks present. Using this approach, we generate 200 unique target modes (see Supplemental Information).

\subsection{Training the CNNs}
In contrast to our previous work~\cite{mastrigt2022machine}, we now ask a neural network to predict the number of intensive modes $b$. We train our networks on $5\times 5$ designs obtained from Monte Carlo sampling of the space. Before training, we preprocess the data.

Because our designs are spatially structured and local interactions between building blocks drive compatible deformations, CNNs are well-suited for predicting the number of intensive modes $b$. We transform our designs into pixelated black-and-white images as input for our CNNs, facilitating straightforward identification of neighboring building blocks capable of deforming compatibly (see Supplemental Information).
Additionally, we use periodic padding to add an extra one-pixel-wide layer to the designs. This allows the network to capture the interactions between building blocks with periodic boundary conditions.

To train the CNNs to predict the number of intensive modes $b_{CNN}$, we generate a set of unit cell designs $\mathbf{X} = \{X_i \}$ and their respective number of intensive modes $\mathbf{b} = \{ b_i \}$. The combination of these designs and their corresponding numbers of intensive modes is referred to as the training set $\mathbf{D}_t = (\mathbf{X}, \mathbf{b})$. The generated data is then divided  into a training (85\%) and a test (15\%) set. Specifically, the training set contains 848898 samples and the set set contains 149982 samples. Our CNNs are trained on imbalanced datasets---designs with a large $b$ become increasingly hard to find using Monte Carlo sampling [Fig.~\ref{fig:b characterization}(A)]. Both the training and test set follow the same distribution of the number of intensive modes $b$; the majority of samples have $b=1$, while there are only 4 samples with $b=6$ in the training set.


We train the CNN to minimize the mean squared error (MSE) between the CNN's prediction $b_{CNN}$ and the true $b$ using the Adam optimization algorithm~\cite{kingma2017adam}. We use 10-fold cross-validation to validate the robustness of our network architecture and training procedure. The network with the lowest MSE over the validation set is selected to be the primary network to use. Specifically, we use a learning rate of $0.0005$, train the network for 100 epochs, and use a batch size of 256. Additionally we use L2-regularization on the weights and biases to reduce overfitting.

\subsection{Neural network architecture\label{sec: NN architecture}}
The CNNs used in this work are composed of three convolution layers for feature extraction, which are connected to a fully-connected hidden layer, which in turn is connected to a single-neuron output layer. 
Specifically, the first convolution layer consists of twenty $2\times 2$ filters with a stride of $(2, 2)$. As the convolution starts in the upper left corner of the input image, the network convolves only $2\times 2$ \textit{plaquettes} between four building blocks. Each plaquette contains a black pixel if one of the four constituent building blocks is oriented such that it has its diagonal interior angle within that plaquette (see Supplemental Information). As such, each plaquette contains information on which building blocks share adjacent diagonal corners and allow for possible compatible deformations of those corners. We conjecture that restricting the network to see only these plaquettes helps achieve a better and more robust performance. 

The second and third convolution layers have 80 and 160 $2\times 2$ filters, respectively,  with a stride of $(1, 1)$. After each convolution operation, we add a trainable bias vector and apply a ReLu activation function on each element of the convolved images. Note that we do not use pooling operations in-between convolution layers.  After the third convolution layer, we flatten the convolved images and fully-connect this vector to a hidden layer of 1000 neurons. Again we add a bias vector and apply the ReLu activation function. The final layer consists of only  a single neuron, and we do not apply an activation function. We take the single output neuron to be $b_{CNN}$, which we aim to be as close to the true $b$ of any input design as possible. We use Jax~\cite{jax2018github} to code our networks.

\subsection{Genetic algorithm}
We employ a genetic algorithm (GA) to explore our CNN beyond its training scope and to prospect highly pluripotent designs. Specifically, we utilize a GA where the fitness function [\eqref{eq: fitness}] is estimated by a trained CNN. The goal of the GA is to achieve a target number of intensive zero modes $b_T$. To achieve this goal, the GA iteratively generates a population of designs (the generation) in three steps: (i) sampling, (ii) fitness evaluation, and (iii) update. Below, we give an overview of these three steps.

In the first step, a fixed number of candidate designs is randomly drawn from the discrete design space. In GA terminology, this set of designs (population) is referred to as generation 0. Only the very first generation is generated in this manner. Second, we score and rank the designs based on their fitness $f$ [\eqref{eq: fitness}]. This fitness is maximal when the CNN's prediction is equivalent to the target number of intensive modes $b_T$. Finally, we use the ranking of designs based on the fitness to generate a new population of designs---the next generation. To efficiently explore the design space, the GA combines and mutates designs. To this end, we employ several standard GA techniques. We select a group of designs from the initial population based on a three-way random tournament selection, and we always include the design with the highest fitness (elitism). This group of designs forms the ``parents''. From this set of parents, we combine designs using a custom crossbreeding scheme (see supplemental information), and we clone to create an additional set of designs. This set of designs also undergoes random mutations of building blocks and produces  the ``children''. The combination of parents and children then forms the next generation, maintaining  the same size as the previous generation. This procedure of fitness evaluation and generation of a new population of designs is repeated until the fitness is maximized or a predefined criterion is met. See the Supplemental Information for more details on the GA.

In short, the GA is able to combine designs to generate new designs with an average $b$, after which random mutations eventually allow the GA to reach a high $b$. However, generating high $b$ designs with the GA requires a significant number of calls  to the evaluation function. Without a fast approximate evaluation function, such as a neural network, generating high $b$ designs would be  unfeasible (see Supplemental Information).

\subsection{Combining and selecting designs for target deformations}
Here, we describe our method to combine and select from a set of $5\times 5$ designs to form larger metamaterials that deform close to desired target deformations. We start from a set of designs generated using our design method and compute the mode structures. We aim to throw away edge modes. We do this by checking the location of CRS sites; if the entire $4\times 4$ inner square of the $5\times 5$ mode consists of CRS blocks, we assume it is an edge mode and disregard it. 

Next, we determine which modes are \emph{compatible} with each other. Here, compatible means that there exists a design that can feature both modes. Modes are compatible if (1) there are no overlapping D sites or (2) if for all overlapping D sites, the building block orientations from the original designs are the same. This means that all modes that originate from the same design are compatible, as they should be. Surprisingly, modes that originate from different designs can also be compatible. By taking the building block orientations of the D sites for both modes, we can create a design that features both modes, as the orientation of CRS blocks is irrelevant. This procedure yields a set of modes and pairs of modes that we label compatible. We represent this as a graph with nodes (modes) and undirected edges (compatible).

To explore our set of modes for combinations that deform close to a target deformation, we search the space of maximal cliques. Such cliques hold the largest set of compatible modes, thus providing the design with the most deformational freedom. We calculate the set of maximal cliques using the Bron-Kerbosch algorithm with vertex ordering by first calculating the degeneracy ordering of the graph~\cite{bron1973algorithm, eppstein2010listing} . For any given set of designs, we  need to perform this calculation only once and save the graph and maximal clique for subsequent searches for target deformations.

To determine if a maximal clique of modes deforms closely to target deformations, we calculate the capacity $C$ [\eqref{eq: C set of M}. This score is calculated using a constraint programming solver that uses satisfiablity methods~\cite{cpsatlp} as described in Materials and Methods C.

To design larger metamaterials, such as the $10 \times 10$ designs of Fig.~\ref{fig:smiley experimental}(B) and \ref{fig:smiley experimental}(D), we split the larger target deformation into local $5\times 5$ deformations. We rank all maximal cliques based on their cumulative capacity $C$ for each $5\times 5$ subset. Then, we take a greedy search approach and combine designs with the highest capacity for each $5\times 5$ subset and calculate the total capacity for the combined design. If the total score in unsatisfactory (below a predefined threshold), we iteratively try the next best designs for each $5\times 5$ subset until we find a satisfactory total score.

\subsection{Prohibiting undesired zero modes}
To prohibit undesired zero modes, we strategically introduce additional rigid bars to the metamaterial structure. To determine the placement of these bars, we analyze the modal structure of the desired and undesired zero modes. In particular, if the undesired mode contains a D block where the desired modes feature CRS blocks, we add a rigid bar across the diagonal to transform the pentodal building block shape to a square shape. This prohibits the building block from deforming with the D mode, effectively prohibiting the undesired mode without introducing  any new (undesired) zero modes. 

\section{Supporting Information}
In this Supporting Information we provide an extended description of our metamaterial, convolutional neural networks (CNNs) and genetic algorithm (GA). Additionally, we provide details on obtaining and preprocessing the training data for our CNNs. Moreover, we show that the combination of local and nonlocal exploration allows our GA to efficiently explore the design space.

\subsection{The metamaterial}
\begin{figure}[b]
    \centering
    \includegraphics{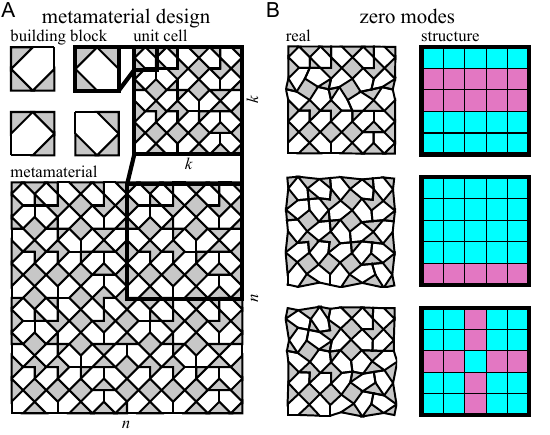}
    \caption{Metamaterial design and zero modes.
    (\textbf{A}) Building blocks (top left) combine into a $k=5$ unit cell (top right) to form a $n=2$ metamaterial (bottom).
    (\textbf{B}) Examples of the three types of zero mode structures: (i) strip-modes (top); (ii) edge-modes (middle); (iii) global modes (bottom).}
    \label{fig:design_modestructure}
\end{figure}

In this work we focus on a combinatorial metamaterial built by tiling building blocks to form $k\times k$ unit cells which are periodically repeated to tile a larger $n\times n$ metamaterial [Fig.~\ref{fig:design_modestructure}(A)]. This metamaterial is composed of a collection of rigid bars and hinges. The building block features two zero modes---infinitesimal deformations that do not stretch any of the bonds up to first order---that we label $m_{CRS}$ and $m_{D}(c)$, where $c$ is the orientation of the building block~\cite{mastrigt2023emergent}.

In previous work, we have showed that there are limitations on the structure of zero modes~\cite{mastrigt2023emergent}. Most importantly, we have shown that areas of adjacent CRS blocks must be rectangular of shape, i.e., their boundaries feature only convex corners. This constraint strongly limits the possible mode structures in our metamaterial. In particular, we distinguish between three types of zero modes: (i) strip-modes, (ii) edge-modes, and (iii) global modes [Fig.~\ref{fig:design_modestructure}(B)]. Each of these mode-types are defined by the spatial ordering of CRS and D blocks. Specifically, strip-modes feature a horizontal or vertical strip of D blocks that spans the entire material sandwiched between two areas of CRS. Edge-modes feature a strip of D blocks that borders the edge(s) of the material, the bulk are CRS blocks. Global modes feature D blocks throughout the entire material.

Additionally, these types of zero modes correspond to a change in the scaling of the number of non-trivial zero modes $N_M(n) = a n + b$. Note that we exclude the three trivial zero modes corresponding to translation and rotation of rigid bodies in two dimensions. Strip-modes are translationally invariant in one direction, resulting in a linear increase of $N_M$ and a contribution to the slope $a$. In contrast, edge-modes and global-modes are not translationally invariant: such modes correspond to an offset of $N_M$ and thus contribute to the number of intensive modes $b$.  

\subsubsection{Statical determinacy}
Generally, the structural integrity of a frame of $N$ joints and $N_B$ rigid bars is captured in the Maxwell-Calladine index theorem in two dimensions~\cite{maxwell1864calculation, calladine1978buckminster}:
\begin{equation}
    N_M - N_{S} = d N - N_{B} - 3 \equiv \mathcal{P},
    \label{eq: maxwell counting}
\end{equation}
where $N_M$ and $N_{S}$ are the number of non-trivial zero modes and states of self-stress respectively. Because $N_S$ is non-negative, the Maxwell count $\mathcal{P}$ is a lower bound for the number of modes, i.e. $N_M\geq \mathcal{P}$. 
Counter-intuitively, frames with intensive modes must be hyperstatic ($\mathcal{P}<0$) for a sufficiently large periodic tiling of the unit cell frame. To see this, we note that for an $n\times n$ tiling of the unit cell frame, $\mathcal{P}(n)$ is a polynomial in $n$ of maximum degree 2. 
To yield an intensive number of modes $N_M$ we require the number of states of self-stress $N_{S}(n)$ to cancel all but the constant term of $\mathcal{P}(n)$. We note that states of self-stress are always localized with open boundary conditions, such that $N_S(n) \propto n^2$ in two dimensions. Thus, we require $\mathcal{P}(n) \propto -n^2$ to yield oligomodal frames.

To see if our metamaterial satisfies this condition, we determine the metamaterial's Maxwell count:
\begin{equation}
    \mathcal{P} = -(n k)^2 + 4 n k + 2
\end{equation}
for a $n\times n$ tiling of a $k\times k$ unit cell. These tilings are strictly hyperstatic for $n k> 2+\sqrt{6} \approx 4.4$. In our paper, where we focus on unit cells of size $k=5$, every tiling is hyperstatic. Thus, to yield intensive modes, the number of states of self-stress should scale as $N_S \propto (n k)^2 - 4 n k$ and cancel the quadratic and linear terms of $\mathcal{P}$. We find that $N_M=a n + b$, thus the quadratic term always cancels. The linear scaling and offset of the number of states of self-stress $N_{S}$ then determines the exact values of $a$ and $b$ in $N_M(n)$ and depends on the orientations of the building blocks.

\subsubsection{Global target modes}
\begin{figure*}[b]
    \centering
    \includegraphics{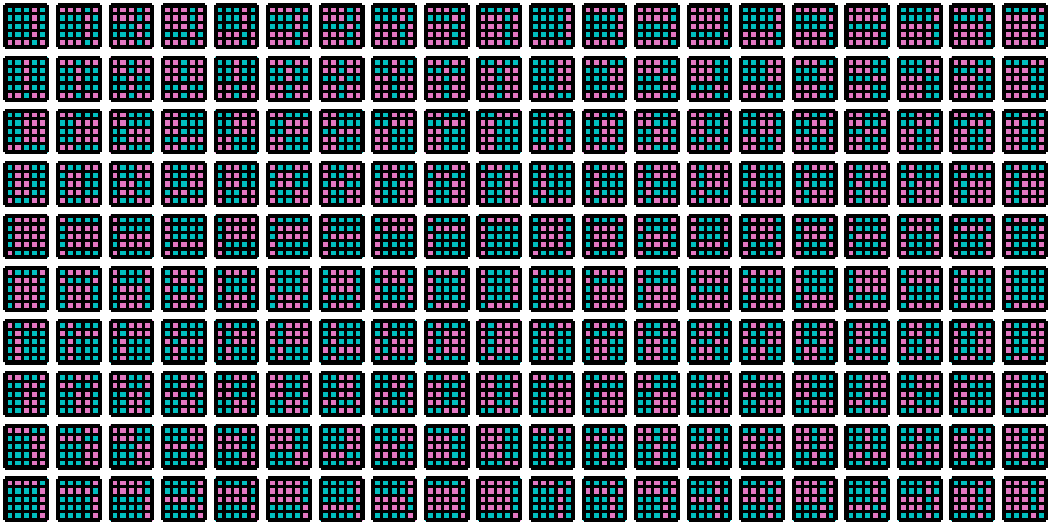}
    \caption{All 200 possible global target deformations generated using the method as described in the Materials and Methods.}
    \label{fig:targets}
\end{figure*}

The 200 possible target deformations are shown in Fig.~\ref{fig:targets}.

\subsubsection{Distribution of edge-modes and global modes.}
\begin{figure}
    \centering
    \includegraphics[width=\columnwidth]{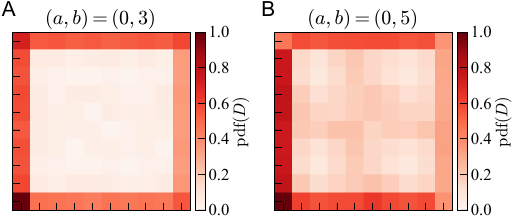}
    \caption{Spatial distribution of the probility density function (pdf) of D blocks for $2\times 2$ tilings of random $5\times 5$ unit cells with $a=0$ extensive modes and $b=3$ (A) or $b=5$ (B) intensive modes.}
    \label{fig:D heatmaps}
\end{figure}

The codimension of the subspace of designs with $b$ intensive modes appears multimodal in distribution, especially for larger $b$. This is most likely due to the different types of modes that contribute to $b$: edge-modes and global modes. For low $b$, most intensive modes are edge-modes [Fig.~\ref{fig:D heatmaps}(A)]. As $b$ increases, global modes become more prevalent [Fig.~\ref{fig:D heatmaps}(B)]. The reason for this is twofold. First, edge-modes consisting of a single line of D blocks require less building blocks with specific orientations than global modes that contain more D blocks (recall that the CRS mode is independent of the building block orientation). Thus, random unit cells are more likely to contain edge-modes than global modes. Second, the number of edge-modes is limited by the number of edges of the metamaterial. Thus, for sufficiently large number of intensive modes $b$ there is a higher probability of global modes. The codimension is related to the probability to prohibit an intensive mode by changing the orientation of a randomly selected building block. The type of mode influences this probability. In general, modes with more D blocks are easier to prohibit as the D mode is sensitive to the orientation of the building block. Additionally, building blocks at the edge are less kinematically restricted than building blocks in the bulk~\cite{mastrigt2023emergent}, making them more robust to changes of orientation.

\subsection{Convolutional neural networks}

\begin{figure*}
    \centering
    \includegraphics{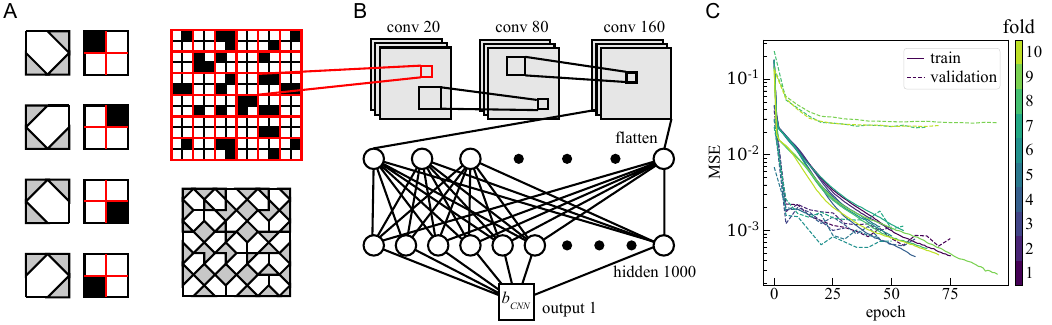}
    \caption{Convolutional neural network (CNN) for metamaterial prediction. 
    (\textbf{A}) To feed our metamaterial designs into a neural network, we represent our designs as a black-and-white image. Building blocks are represented as $2\times 2$ plaquettes of pixels, one black and three white (left). $5\times 5$ designs (bottom right) translate to $10\times 10$ pixel images. Additionally, we pad these images using periodic boundary conditions with one additional layer of pixels, so that we end up with a $12\times 12$ pixel image (top right).
    (\textbf{B}) The pixel image of (A) forms the input for our CNN, which consists of three convolutional layers, a single hidden layer and a single output node.
    (\textbf{C}) The training (solid line) and validation (dashed line) mean squared error (MSE) over the training epochs for each fold (colorbar) in our 10-fold cross-validation. Note that we used an early stopping condition to prevent overfitting, resulting in different number of training epochs for the folds.}
    \label{fig:pixel_CNN_train}
\end{figure*}

We use 10-fold cross-validation to validate the robustness of our network architecture and training procedure. The network with the lowest mean squared error (MSE) over the validation set is selected to be the primary network to use. Specifically, we use a learning rate of $0.0005$, train the network for 100 epochs and use a batch size of 256. Additionally we use L2-regularization on the weights and biases. The architecture of our CNN is shown schematically in Fig.~\ref{fig:pixel_CNN_train}(B). The training process for each fold is shown in Fig.~\ref{fig:pixel_CNN_train}(C).

\subsubsection{Preprocessing \label{sec: preprocessing}}
The configurational data of the unit cell designs is preprocessed to a pixelated representation (see Fig.~\ref{fig:pixel_CNN_train}(A)) for input to the neural network. We found that representing the unit cell designs in this pixelated representation improved convergence during training and better performance of the trained networks over the validation sets. We believe that this is due to the orientations of the building blocks being clearly represented visually as opposed to simply representing the orientations as integers in a matrix. Because each building block is represented as a $2\times 2$ square, we can choose where the convolutional layer applies its filters more finely. As discussed in the Methods and Materials section, this allows us to let the networks `see' only the interactions between building blocks. Additionally, we use periodic padding to pad the designs with an extra one pixel wide layer to allow the network to see interactions between building blocks with periodic boundary conditions.



\subsection{Genetic algorithm}
\begin{figure*}
    \centering
    \includegraphics{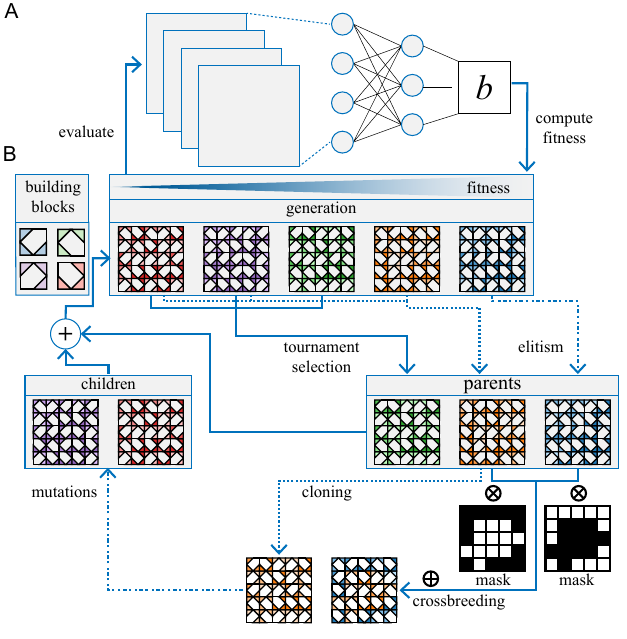}
    \caption{Schematic representation of our genetic algorithm.
    (\textbf{A}) A CNN calculates the number of intensive modes $b_{CNN}$ for each design in the generation.
    (\textbf{B}) A generation of designs (top) is ranked based on the fitness $f$. Using a three-way tournament selection the algorithm selects designs from the generation that, together with the fittest design in the generation, form the parents (right). From the parents, we generate new designs in two ways: cloning and crossbreeding (bottom). In cloning, we simply make a copy of the parent design chosen at random. In crossbreeding, we randomly select two parents that we combine using two-dimensional masks (black-and-white matrices). Additionally, building blocks in these newly generated designs have the chance to mutate into new orientations. The mutated designs form the children (left) who, together with the parents, form the next generation. This iterative procedure continues until a sufficiently high fitness or stopping condition is reached.}
    \label{fig:GA architecture}
\end{figure*}

To explore the discrete design space, we use a genetic algorithm (see Fig.~\ref{fig:GA architecture} for a schematic overview). The general idea of such an algorithm is to find suitable candidates, specifically with a high ``fitness'', by selection and combination of the most fit candidates in a population to create a new population of candidates: this is the new generation. As genetic algorithms are known to go through a lot of generations, the computational bottleneck of such algorithms is typically the calculation of the fitness for each new generation. To save computational time, we use a CNN to calculate the fitness $f$. Without such an approximate fitness function, the computational time increases by a factor 390 and exploration of the design space is unfeasible.

Specifically, our genetic algorithm has a population size of 100, of which 29 are selected to be parents using a three-way tournament selection. Additionally, the fittest candidate in the population is automatically selected to be one of the parents, so that we have a total of 30 parents. To create a new generation of 100 designs, we require 70 children. Of those 70 children, half are created using cross-breeding of randomly selected pairs of parents. To combine the design of two parents to create a new child, we use a Gaussian filter to filter a random binary $5\times 5$ mask, which we then again binarize to create a mask where half of the elements are 0 and half are 1. This mask is then used to take part of the two parents and combine them to create a new child [Fig.~\ref{fig:GA architecture}(A)]. The reason we use this method over more standard methods, such as k-point crossover or uniform crossover, is that we believe local clusters of building blocks are important for intensive zero modes. The other 35 children are taken by cloning randomly selected parents. All building blocks in the children designs have a chance of 10\% to mutate to a different orientation. Each of these different orientations are equally likely to be selected. The combination of the parents and children after mutation form the new generation of designs. 

\subsubsection{GA exploits nonlocal structure}

To understand how the GA explores design space, we investigate the two key exploration techniques of our GA. 
Underlying the evolution of generations of designs are two main processes: (i) cloning and (ii) crossbreeding [Fig.~\ref{fig:GA architecture}(B)]. Our GA thus explores the design space in two ways: (i) the cloned designs undergo local mutations. This is local exploration of the design space surrounding the cloned parent design. (ii) The crossbred designs are combinations of two parent designs on top of local mutations. This is non-local exploration of the design space.

\begin{figure}[b!]
    \centering
    \includegraphics[width=\columnwidth]{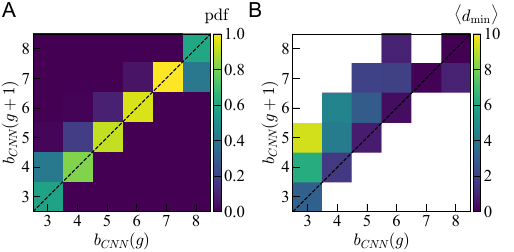}
    \caption{GA design space exploration.
    (\textbf{A}) Pdf for the design with the highest fitness in a GA run to transition from $b_{CNN}(g)$ to $b_{CNN}(g+1)$ for iterative generations. For low $b_{CNN}(g)$ the GA is more likely to transition to a higher $b_{CNN}$. For high $b_{CNN}(g)$ the GA struggles to quickly find designs with a larger $b_{CNN}(g+1)$. Note that $b_{CNN}(g)$ can never transition to a lower $b_{CNN}(g+1)$ as the GA always takes the design with the highest $b_{CNN}$ to the next generation.
    (\textbf{B}) The average minimal distance $\langle d_{min} \rangle$, where $\langle . \rangle$ denotes the average over an ensemble of GA runs, for the fittest design in generation $g+1$ with $b_{CNN}(g+1)$ as a function of $b_{CNN}(g)$ of the fittest design in generation $g$. As $b_{CNN}(g)$ increases, the distance $\langle d_{min} \rangle$ between the fittest design in the next generation $g+1$ and all designs in generation $g$ decreases.}
    \label{fig:GA exploration}
\end{figure}

The combination of both local and non-local exploration are crucial for the success of the GA.
To illustrate this, we determine the minimal distance between between the design $X$ with the highest fitness of generation $g+1$ and every other design $Y$ in generation $g$ as
\begin{equation}
     d_{min} = \mathrm{min}_{\{Y\}}\left( \sum_{i, j} 1 - \delta(X_{i, j}, Y_{i, j}) \right),
    \label{eq: distance}
\end{equation}
where $\delta(x, y)$ is the Kronecker delta function and $X_{i, j}$ is the orientation of the building block at site $(i, j)$ in design $X$. \eqref{eq: distance} thus captures the number of different building blocks between design $X$ of generation $g+1$ and the design $Y$ in generation $g$ that is closest to $X$. Non-local exploration primarily plays a role for early generations when $b_{CNN}(g)$ is likely to be small, which allows the GA to find designs of a higher $b_{CNN}$ (Fig.~\ref{fig:GA exploration}(C)). For larger $b_{CNN}$, in later generations, local exploration is key. 

Intuitively, the GA is able to efficiently explore the design space by first crossbreeding designs to quickly find designs with a reasonably high number of intensive modes $b_{CNN}$. This is most likely a consequence of the fact that most designs with a low number of intensive modes feature deformations that are localized on the edge of the material. Such edge modes are less sensitive to building block mutations than global modes and thus crossbreeding is more likely to combine designs that feature edge modes. Moreover, as the number of intensive modes $b$ increases the probability to inhibit any mode by changing orientations increases, such that the crossbreeding is more likely to result in a net conservation or decrease of $b$.
After $b=5$, the GA appears to rely on cloning to locally explore around an ensemble of designs with high $b$ to find rare $b+1$ designs. 
Thus, the GA is able to efficiently generate designs that feature a large number of intensive modes $b$ thanks to both non-local and local exploration, resulting in successful runs that generally converge.



\subsubsection{Random walks}
To both obtain starting designs for the random walks and compare the evolution of GA designs, we perform and keep track of a hundred GA runs with $b_{T}=7$. We start a hundred random walks from the final hundred designs with the highest fitness, thus the starting $b(s=0)$ varies from $5\leq b \leq 7$.

\subsubsection{Comparison to other methods.}
\begin{figure}
    \centering
    \includegraphics{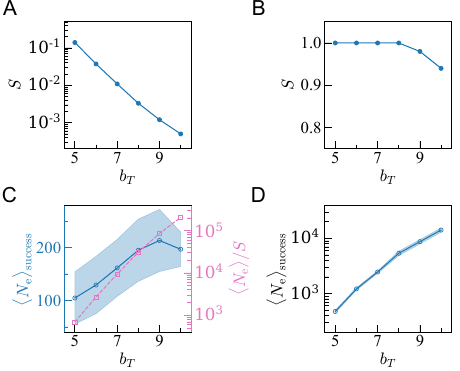}
    \caption{Comparison of hill-climbing and genetic algorithm. 
    (\textbf{A}) Success rate $S$ of 10000 hill-climbing runs to reach target number of intensive modes $b_T$ decreases exponentially. 
    (\textbf{B}) The success rate $S$ of 10000 GA-runs to reach $b_T$.
    (\textbf{C}) The average (blue circles) and standard deviation (blue shaded area) of the number of evaluations of the fitness function $f$ for successful runs $\langle N_e \rangle_{\mathrm{success}}$ to reach $b_T$ using hill-climbing. The average number of evaluations per successive run $\langle N_e \rangle \ S$ (pink squares) increases exponentially with $b_T$.
    (\textbf{D}) The average (blue circles) and standard deviation (shaded area) of the number of evaluations of the fitness function $f$ for successful runs $\langle N_e \rangle_{\mathrm{success}}$ increases exponentially with $b_T$.}
    \label{fig:HC vs GA}
\end{figure}

Our GA is able to find ultra-rare designs with a large number of intensive modes $b$, that is not possible using random search as designs with high $b$ become exponentially more rare with $b$. Alternatively, one could try to exploit the hierarchical design space structure through a hill-climbing method. While this approach succeeds sometimes, it fails an exponentially larger fraction of the time for increasing target number of intensive modes $b_T$ [Fig.~\ref{fig:HC vs GA}(A)-(B)]. This results in a larger average number of evaluations of the fitness function $f$ for large $b_T$ to make a successful run [Fig.~\ref{fig:HC vs GA}(C)-(D)]. Compared to state-of-the-art generative methods, such as variational autoencoders (VAE), generative adversarial neural networks (GANs), and normalizing flows, our method allows for extrapolation outside the scope of the training set. These generative methods all aim to approximate the (unknown) underlying probability distributions of the training set---without examples such methods are unable to generate designs with the desired property.

\bibliography{bibliography}
\end{document}